\documentclass[sigconf]{acmart}

\pdfoutput=1
\usepackage{nopageno}
\usepackage{makecell}
\usepackage{subcaption}
\usepackage{caption}
\usepackage{kantlipsum}
\usepackage{float}
\usepackage{booktabs} 
\usepackage{draftwatermark}
\usepackage{balance}
\usepackage[keeplastbox]{flushend}
\SetWatermarkText{}

\setcopyright{none}

\acmConference[MLDM4P'18]{2018 KDD Workshop on Machine Learning and Data Mining for Podcasts}{August 2018}{
  London, UK}
\acmYear{2018}
\acmISBN{123-4567-24-567/08/06}
\copyrightyear{2018}
\acmDOI{}
\makeatletter
\renewcommand\@formatdoi[0]{\ignorespaces}
\makeatother
\begin{document}

\title{Recognizing Film Entities in Podcasts}

\author{Ahmet Salih Gundogdu}

\affiliation{%
  \institution{WanerMedia Applied Analytics}
  \city{Boston}
  \state{MA}
  \postcode{02116}
}
\email{ahmet.gundogdu@turner.com}

\author{Arjun Sanghvi}

\affiliation{%
  \institution{WanerMedia Applied Analytics}
  \city{Boston}
  \state{MA}
  \postcode{02116}
}
\email{arjun.sanghvi@turner.com}

\author{Keith Harrigian}

\affiliation{%
  \institution{WanerMedia Applied Analytics}
  \city{Boston}
  \state{MA}
  \postcode{02116}
}
\email{keith.harrigian@turner.com}

\renewcommand{\shortauthors}{Gundogdu et al.}

\begin{abstract}
In this paper, we propose a Named Entity Recognition (NER) system to identify film titles in podcast audio. Taking inspiration from NER systems for noisy text in social media, we implement a two-stage approach that is robust to computer transcription errors and does not require significant computational expense to accommodate new film titles/releases. Evaluating on a diverse set of podcasts, we demonstrate more than a 20\% increase in F1 score across three baseline approaches when combining fuzzy-matching with a linear model aware of film-specific metadata.
\end{abstract}




\maketitle

\section{Introduction}
\thispagestyle{empty}
Podcast audiences have more than doubled in size over the last decade, bringing with them demand for more frequent releases and an expanded scope of content \cite{WinNT}. In parallel, the barrier to entry for hosting a podcast has been significantly lowered, with reduced costs for audio recording technology allowing non-experts to engage in discussion of topics such as politics or entertainment that were previously reserved for more traditionally-accredited and financially-supported individuals \cite{baumgartner2010myfacetube, kang2012differences}. In this combination, podcasts have demonstrated tremendous potential as a gauge for social opinion.

Nonetheless, the volume of podcast content produced today necessarily precludes large-scale topical analysis. In particular, it is currently difficult to track mentions of noteworthy properties across multiple podcast channels. As such, NER, the identification of qualitatively significant word phrases such as people and organizations \cite{chieu2002named}, is a well-needed focus in the area of podcast analysis. NER is a key step necessary to make higher level inferences such as measuring sentiment, identifying emotions associated with properties, or building predictive models for property-level response variables such as revenue based on podcast data. 

Although NER systems for formal, written language perform accurately \cite{lin2009phrase, ratinov2009design}, there remains substantial room for improvement in evolving communication mediums where traditional linguistic structures are used less consistently (e.g. social media posts and informal conversations). Entity recognition faces further challenges in the domain of human speech, where an intermediary step to transform audio into computer-readable text entirely removes orthographic features, which are often used to highlight entities in writing (e.g. captitalization, punctuation).

To address the aforementioned challenges for entity detection in podcasts, we propose a two-stage NER system and evaluate it in the context of film title detection. We note that detection of film titles is one of the more niche and ambitious tasks within this research domain, given that new films are released each week and most NER systems rely on large volumes of training data that can be costly to obtain. Given this scope, we believe that a successful film title detection approach has promise to transfer to other entity classes which are traditionally more stable over time.

In this paper, we begin by briefly reviewing existing NER systems used in informal language domains such as human speech, highlighting their limitations in the context of film title detection. In Section 3, we discuss our data sources, external dependencies and our pre-processing approach. In Section 4, we propose and detail the two-step candidate identification and entity classification procedure that lies at the crux of this paper. Finally, in Section 5, we evaluate our proposed method and compare it against three baselines.

\section{Related Work}
Although there exists plenty of research on NER for traditional noun phrases such as people, locations, and organization names, little has been done for niche entities such as movies and books. The challenge in the latter is that these properties evolve in much shorter time intervals (i.e. new movies are released every week).

Prior research has used supervised machine learning to recognize entities from audio using acoustic features. For example, the speech recognition model proposed by \cite{hatmi2013incorporating} uses constrained maximum likelihood linear regression to simultaneously predict the most probable sequence of words and entity classes within an audio waveform. However, this approach requires a large set of token-labeled and time-synchronized training transcripts. Such a dependency is prohibitive in the dynamically changing landscape for film titles.

Focusing less on acoustic signal, \cite{chowdhury2013simple} feeds word tokens and linguistic features from audio transcripts into a Conditional Random Field (CRF) model to detect people, locations, organizations, and geo-political entities. While this approach works quite well, it is not suited for unstable entity classes and proves quite vulnerable to word/phrase transcription errors. 

Given the limitations of previous research in the audio analysis domain, we look for inspiration in another challenging medium - social media. Notably, language in social media typically suffers from inconsistent spelling, poor grammar, and a dearth of orthographic features \cite{eisenstein2013bad}. Given the current state of automatic audio transcription tools, the computer-readable text generated from podcast audio often encounters the same inconsistencies. Moreover, language in both social media and human speech requires normalization, as people often invoke non-standard tokens and phrases within these domains \cite{liu2012broad}.

In a relatively recent study, \cite{Ashwini} propose a system that identifies entity candidates in social media by matching token sequences in tweets to phrases in a gazetteer.  To address issues with precision, they train a classifier using a combination of orthographic, n-gram, and syntactic features to determine whether entity candidates are indeed true entities. Importantly, their system does not require constant re-training, as the gazetteer may be updated with new elements in real time. While their results demonstrate promise for a variety of entity types, their system still relies on capitalization, special characters, and syntactic features to achieve the desired performance. As mentioned above, these features are critically absent in transcripts of podcast audio and thus motivate our work. 

\section{Data and Pre-processing}
We collect 20 film-related podcasts from various publicly available channels (see Table 1), including National Public Radio, SlashFilm, Screen Junkies and Looper. We listen to each podcast and manually note film properties mentioned within each one to serve as the ground truth. The quality of labels is evaluated by the Cohen's Kappa inter-annotator agreement (0.63). The podcasts are of similar length (10 $\pm{5}$ minutes) except those from SlashFilm (100 minutes). The complete distribution of entities within our dataset and estimated transcription errors can be found in Table 1.

\begin{table}[]
\centering
\caption{Podcast Label Distribution and  Average Word Error Rates}
\begin{tabular}{@{}lllllll@{}}

\toprule
{Channels} & {\#1} & {\#2} & {\#3} & {\#4} & {Total} & {\begin{math}\overline{WER}\end{math}} \\ \midrule
Collider                       & 51                             & 2                              & -                              & -                              & 53                         & 2\%                      \\
All Things Consd.              & 4                              & 2                              & -                              & -                              & 6                          & 26\%                     \\
Looper                         & 4                              & 16                             & 37                             & -                              & 57                         & 2\%                      \\
Bob Mondello                   & 20                             & 15                             & 5                              & -                              & 40                         & 10\%                     \\
Morning Ed.                    & 3                              & 11                             & -                              & -                              & 14                         & 23\%                     \\
Screen Junkies                 & 20                             & 30                             & 3                              & 11                             & 64                         & 2.5\%                    \\
Angry Joeshow                  & 18                             & -                              & -                              & -                              & 18                         & 4\%                      \\
Fresh Air                      & 14                             & 12                             & -                              & -                              & 26                         & 25\%                     \\
SlashFilm                          & 38                             & -                              & -                              & -                              & 38                         & 38\%                    
\end{tabular}
\vspace{-1.0em}
\end{table}

Each podcast is subjected to the same set of pre-processing steps. First, raw podcast audio is transcribed using an open-source speech recognition framework from \cite{Zhang}. The output is a long sequence of lowercase words separated by whitespace; there are not any orthographic features or punctuation, which notably have been deemed critical in existing NER systems. Applying \cite{Zhang}'s model to audio from a National Public Radio podcast in which a human-curated transcript is available, we empirically observe a 23\% Word Error Rate (WER) on average. Given this relatively high WER, we reiterate the importance of an entity-detection system that is flexible enough to handle errors in the transcription procedure.

To aid in the inference of syntactic features, punctuation is inferred using a bidirectional recurrent neural network (with attention) that has been trained on European Parliment speech data \cite{tilk2016bidirectional}. To evaluate the quality of this inference, we apply it to the same podcast used to test WER. The punctuation model performs significantly better than chance and achieves a precision, recall, and F1 score of 0.78, 0.64, and 0.70, respectively.

After punctuation is complete, all numeric values are converted to text (e.g. 1984 to nineteen eighty four). Finally, the processed text string is tokenized using the sentence and word tokenizers from NLTK \cite{bird2009natural}.

\section{Methods}
\subsection{Identifying Candidates}
\subsubsection{Gazetteer}
We use a proprietary database of 9000 films produced between 2000 and 2016 as our gazetteer. Each film has the following metadata available: production budget, keywords, plot summary, and logline. Notably, 70\% of films are missing data from at least one of these fields. 

\subsubsection{Entity Lookup}
\cite{Ashwini} performs exact string matching using suffix trees to identify words and phrases which may be entities (a.k.a. entity candidates). However, exact string matching tends to miss titles that have minor word errors (e.g. film "Coco" transcribed as "cocoa"). In our approach, we use a Levenshtein ratio \cite{fuad2012towards} with a threshold that varies with the number of tokens present in the phrase to determine whether a match exists. The threshold for each n-gram length  \begin{math}n \in [1,6]\end{math} is determined using cross-validation within the training set. This similarity measure is implemented to account for the relatively high WER of the \cite{Zhang} audio transcription model. Accordingly, long movie titles such as "Three Billboards Outside Ebbing, Missouri" can have minor transcription errors without being ignored during the entity candidate identification stage. 

\begin{figure*}[h]
\centering
\includegraphics[scale=0.45]{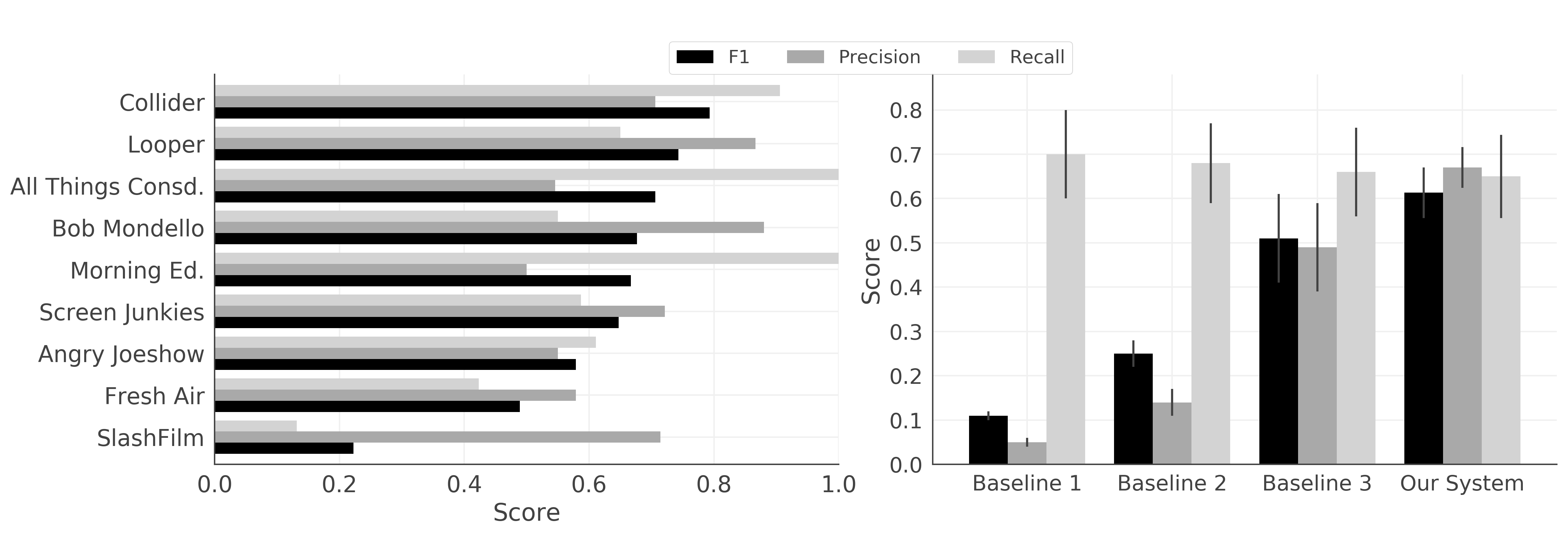}
\caption{Left: Results from LOCO cross-validation. Variation due to the differences in WER and total number of true film title mentions. Right: Comparison of our NER system to 3 baselines across LOCO cross-validation. Substantial gain in precision achieved using metadata-based rules and features.}
\end{figure*}

\subsubsection{Feature Extraction}
Using the tokenized text output from the preprocessing stage in Section 3, we infer part of speech (POS) tags for each token in the audio transcript using the Stanford POS Tagger \cite{bird2009natural}. Then, we add several features to each entity candidate based on metadata from their potential film match. All features can be found in Table 2.

\begin{table*}
\centering
  \caption{Feature Properties Extracted from Transcription}
  \label{tab:commands}
  \def\arraystretch{1.5}
  \begin{tabular}{ccl}
    \toprule
    \makecell{Features} &\makecell{Properties} & \makecell{Definition}\\
    \midrule
    \texttt{Closeness} & \begin{math}  Closeness(m,k) = 1 - \frac{|w_m-w_{k}|} {||transcript||} \end{math} & \makecell{Closeness value between movie \begin{math}m\end{math} and keyword \begin{math}k\end{math}} \\
    \texttt{Levenshtein Ratio}& \begin{math}  \frac{lev(m_{matched},m_{original})} {\max(||m_{matched}||,||m_{original}||)} \end{math} & \makecell{\begin{math}lev(a,b)\end{math} measures the minimum number of deletions,\\insertions and substitutions to transform \begin{math}a\end{math} into \begin{math}b\end{math}} \\
    \texttt{Production Budget}& \begin{math}  \frac{budget_m-budget_{min}} {budget_{max}-budget_{min}} \end{math} & \makecell{Normalized budget for film \begin{math}m\end{math} within the transcript}\\
    \texttt{Title POS-tags}& \begin{math} [\textit{POS-tags}]_m  \end{math} & \makecell{Bag of POS-tags in the title of movie \begin{math}m\end{math}}\\
    \texttt{Pre/Post POS-tags}&  \begin{math} [\textit{POS-tag}]_{m-1} \end{math},  \begin{math} [\textit{POS-tag}]_{m+1} \end{math} & \makecell{POS-tags of preceeding and succeeding word for candidate}\\
    \texttt{N-gram Levels}& \begin{math}  [1-6] \end{math} & \makecell{Number of tokens in entity candidate}\\
    \bottomrule
  \end{tabular}
\vspace{-1.5em}
\end{table*}

Notably, we design a metric to capture the thematic relevance the context around an entity candidate has to its associated film. Informally, we define \textit{closeness} to be a normalized value representing the number of words in the transcript which seperate the entity candidate, \begin{math}m\end{math}, from relevant keyword, \begin{math}k\end{math}, where \begin{math}k\in K\end{math} and \begin{math}K\end{math} denotes keywords of the movie \begin{math}m\end{math} in our database. Their word indices within the transcript are noted by \begin{math}w_m\end{math} and \begin{math}w_{k}\end{math} respectively. We define this metric mathematically in Table 1. After calculating each \begin{math}k\end{math}'s closeness value to corresponding entity candidate \begin{math}m\end{math}, we extract \begin{math}\frac{1}{||K||}\sum_{k\in K}closeness({w_m,k})\end{math}, \begin{math}\min_{k\in K}closeness({w_m,k})\end{math} and \begin{math}\max_{k\in K}closeness({w_m,k})\end{math} values as model features. 
\noindent For example, if the entity candidate is "Godzilla", we may expect words like "monster", "large", or "japan" to appear in close proximity to the entity candidate's position within the transcript. 

\subsection{Classifying Entities}
After identifying potential film mentions via our fuzzy matching algorithm, the entity candidates are subjected to binary classification via logistic regression. We evaluated baseline approaches and our system using 9-fold Leave One Channel Out (LOCO) cross-validation. Results from our approach are highlighted in the left-side of Figure 1. Hyperparameters (regularization and penalty) are selected to optimize F1 score within each training set. 

Since the model is agnostic to specific words or phrases, this NER system does not require any retraining when the gazetteer is updated to reflect new movie releases.

\subsubsection{Feature Selection}
We find that the most predictive features of true entity mentions include the following: n-gram levels, POS-tags, closeness, and Levenshtein ratio (using null hypothesis testing p-values<0.05 for mentioned features). While n-gram levels and POS-tags provide adequate performance, the most significant performance gain comes from the addition of the metadata-based features: closeness and production budget.

\section{Results and Discussion}
We compare performance of our model-based entity recognition to three baseline approaches (right-side of Figure 1). Baseline 1 classifies all entity candidates as true film mentions. Baseline 2 is similar to Baseline 1, except that we limit candidates to those inferred to be a noun-phrase using the Stanford POS Tagger. For Baseline 3, we consider all candidates identified in the first stage of our process and then remove those whose closeness statistics are below thresholds determined via cross-validation over the training data.

Although \cite{ritter2011named} has proven useful in the social media domain, we find it does not serve as an adequate baseline for our task. The pre-trained model from \cite{ritter2011named} identifies zero film titles across our dataset of transcripts. To understand why, we applied this model to a human annotated transcription from National Public Radio (NPR) with and without capitalization. While the model identified 16 out of 33 true film title mentions in the capitalized transcript, it did not identify any within the uncapitalized version. 
\noindent The lack of capitalization dramatically reduces the performance and highlights the value of a gazetteer-based approach. 

The rule-based approach (Baseline 3) demonstrates that the metadata adds most of the predictive power. The linear model is better at taking into account multiple features as compared to the rule-based approach. As shown in the right-side of Figure 1, our model achieves an average F1 score, precision and recall of 0.61, 0.67 and 0.65, respectively.

To estimate the effect that the high WER from \cite{Zhang}'s speech transcription model has on our overall method, we also apply our two-stage NER system to the NPR podcast used to evaluate \cite{ritter2011named}'s NER model. We find that our system correctly identifies 27 true film mentions in the human-curated transcript as opposed to 22 true mentions in the computer-generated transcript. As such, we believe our system has room to improve given access to more accurate speech transcription models.

Future research will explore two key directions. First, we plan to include additional film metadata fields such as release date, production studio, and cast members as features in the candidate classification model. We hypothesize that several of these fields can be represented within the model in a similar fashion to keyword mentions. Second, we plan to source a larger and more granularly labeled set of podcast transcript data to allow the use of data-greedy sequence learning models.

\bibliographystyle{ACM-Reference-Format}
\balance
\bibliography{sample-bibliography}


\begin{thebibliography}{00}


\ifx \showCODEN    \undefined \def \showCODEN     #1{\unskip}     \fi
\ifx \showDOI      \undefined \def \showDOI       #1{#1}\fi
\ifx \showISBNx    \undefined \def \showISBNx     #1{\unskip}     \fi
\ifx \showISBNxiii \undefined \def \showISBNxiii  #1{\unskip}     \fi
\ifx \showISSN     \undefined \def \showISSN      #1{\unskip}     \fi
\ifx \showLCCN     \undefined \def \showLCCN      #1{\unskip}     \fi
\ifx \shownote     \undefined \def \shownote      #1{#1}          \fi
\ifx \showarticletitle \undefined \def \showarticletitle #1{#1}   \fi
\ifx \showURL      \undefined \def \showURL       {\relax}        \fi
\providecommand\bibfield[2]{#2}
\providecommand\bibinfo[2]{#2}
\providecommand\natexlab[1]{#1}
\providecommand\showeprint[2][]{arXiv:#2}

\bibitem[\protect\citeauthoryear{??}{Win}{}]%
        {WinNT}
\bibinfo{title}{Audio podcast consumption in the U.S. 2018 | Statistic}.
\newblock
  \bibinfo{howpublished}{\url{https://www.statista.com/statistics/270365/audio-podcast-consumption-in-the-us/}}.
    (\bibinfo{year}{????}).
\newblock
\newblock
\shownote{Retrieved May 7, 2018.}


\bibitem[\protect\citeauthoryear{Ashwini and Choi}{Ashwini and Choi}{2014}]%
        {Ashwini}
\bibfield{author}{\bibinfo{person}{Sandeep Ashwini} {and}
  \bibinfo{person}{Jinho~D. Choi}.} \bibinfo{year}{2014}\natexlab{}.
\newblock \showarticletitle{Targetable Named Entity Recognition in Social
  Media}.
\newblock \bibinfo{journal}{{\em arXiv preprint arXiv:1408.0782\/}}
  (\bibinfo{year}{2014}).
\newblock


\bibitem[\protect\citeauthoryear{Baumgartner and Morris}{Baumgartner and
  Morris}{2010}]%
        {baumgartner2010myfacetube}
\bibfield{author}{\bibinfo{person}{Jody~C Baumgartner} {and}
  \bibinfo{person}{Jonathan~S Morris}.} \bibinfo{year}{2010}\natexlab{}.
\newblock \showarticletitle{MyFaceTube politics: Social networking web sites
  and political engagement of young adults}.
\newblock \bibinfo{journal}{{\em Social Science Computer Review\/}}
  \bibinfo{volume}{28}, \bibinfo{number}{1} (\bibinfo{year}{2010}),
  \bibinfo{pages}{24--44}.
\newblock


\bibitem[\protect\citeauthoryear{Bird, Klein, and Loper}{Bird
  et~al\mbox{.}}{2009}]%
        {bird2009natural}
\bibfield{author}{\bibinfo{person}{Steven Bird}, \bibinfo{person}{Ewan Klein},
  {and} \bibinfo{person}{Edward Loper}.} \bibinfo{year}{2009}\natexlab{}.
\newblock \bibinfo{booktitle}{{\em Natural language processing with Python:
  analyzing text with the natural language toolkit}}.
\newblock \bibinfo{publisher}{" O'Reilly Media, Inc."}.
\newblock


\bibitem[\protect\citeauthoryear{Chieu and Ng}{Chieu and Ng}{2002}]%
        {chieu2002named}
\bibfield{author}{\bibinfo{person}{Hai~Leong Chieu} {and}
  \bibinfo{person}{Hwee~Tou Ng}.} \bibinfo{year}{2002}\natexlab{}.
\newblock \showarticletitle{Named entity recognition: a maximum entropy
  approach using global information}. In \bibinfo{booktitle}{{\em Proceedings
  of the 19th international conference on Computational linguistics-Volume 1}}.
  Association for Computational Linguistics, \bibinfo{pages}{1--7}.
\newblock


\bibitem[\protect\citeauthoryear{Chowdhury}{Chowdhury}{2013}]%
        {chowdhury2013simple}
\bibfield{author}{\bibinfo{person}{Md~Faisal~Mahbub Chowdhury}.}
  \bibinfo{year}{2013}\natexlab{}.
\newblock \showarticletitle{A simple yet effective approach for named entity
  recognition from transcribed broadcast news}.
\newblock In \bibinfo{booktitle}{{\em Evaluation of Natural Language and Speech
  Tools for Italian}}. \bibinfo{publisher}{Springer}, \bibinfo{pages}{98--106}.
\newblock


\bibitem[\protect\citeauthoryear{Eisenstein}{Eisenstein}{2013}]%
        {eisenstein2013bad}
\bibfield{author}{\bibinfo{person}{Jacob Eisenstein}.}
  \bibinfo{year}{2013}\natexlab{}.
\newblock \showarticletitle{What to do about bad language on the internet}. In
  \bibinfo{booktitle}{{\em Proceedings of the 2013 conference of the North
  American Chapter of the association for computational linguistics: Human
  language technologies}}. \bibinfo{pages}{359--369}.
\newblock


\bibitem[\protect\citeauthoryear{Fuad}{Fuad}{2012}]%
        {fuad2012towards}
\bibfield{author}{\bibinfo{person}{Muhammad Marwan~Muhammad Fuad}.}
  \bibinfo{year}{2012}\natexlab{}.
\newblock \showarticletitle{Towards Normalizing the Edit Distance Using a
  Genetic Algorithms--Based Scheme}. In \bibinfo{booktitle}{{\em International
  Conference on Advanced Data Mining and Applications}}. Springer,
  \bibinfo{pages}{477--487}.
\newblock


\bibitem[\protect\citeauthoryear{Hatmi, Jacquin, Morin, and Meigner}{Hatmi
  et~al\mbox{.}}{2013}]%
        {hatmi2013incorporating}
\bibfield{author}{\bibinfo{person}{Mohamed Hatmi}, \bibinfo{person}{Christine
  Jacquin}, \bibinfo{person}{Emmanuel Morin}, {and} \bibinfo{person}{Sylvain
  Meigner}.} \bibinfo{year}{2013}\natexlab{}.
\newblock \showarticletitle{Incorporating named entity recognition into the
  speech transcription process}. In \bibinfo{booktitle}{{\em Proceedings of the
  14th Annual Conference of the International Speech Communication Association
  (Interspeech'13)}}. \bibinfo{pages}{3732--3736}.
\newblock


\bibitem[\protect\citeauthoryear{Kang and Gretzel}{Kang and Gretzel}{2012}]%
        {kang2012differences}
\bibfield{author}{\bibinfo{person}{Myunghwa Kang} {and} \bibinfo{person}{Ulrike
  Gretzel}.} \bibinfo{year}{2012}\natexlab{}.
\newblock \showarticletitle{Differences in social presence perceptions}.
\newblock In \bibinfo{booktitle}{{\em Information and Communication
  Technologies in Tourism 2012}}. \bibinfo{publisher}{Springer},
  \bibinfo{pages}{437--447}.
\newblock


\bibitem[\protect\citeauthoryear{Lin and Wu}{Lin and Wu}{2009}]%
        {lin2009phrase}
\bibfield{author}{\bibinfo{person}{Dekang Lin} {and} \bibinfo{person}{Xiaoyun
  Wu}.} \bibinfo{year}{2009}\natexlab{}.
\newblock \showarticletitle{Phrase clustering for discriminative learning}. In
  \bibinfo{booktitle}{{\em Proceedings of the Joint Conference of the 47th
  Annual Meeting of the ACL and the 4th International Joint Conference on
  Natural Language Processing of the AFNLP: Volume 2-Volume 2}}. Association
  for Computational Linguistics, \bibinfo{pages}{1030--1038}.
\newblock


\bibitem[\protect\citeauthoryear{Liu, Weng, and Jiang}{Liu
  et~al\mbox{.}}{2012}]%
        {liu2012broad}
\bibfield{author}{\bibinfo{person}{Fei Liu}, \bibinfo{person}{Fuliang Weng},
  {and} \bibinfo{person}{Xiao Jiang}.} \bibinfo{year}{2012}\natexlab{}.
\newblock \showarticletitle{A broad-coverage normalization system for social
  media language}. In \bibinfo{booktitle}{{\em Proceedings of the 50th Annual
  Meeting of the Association for Computational Linguistics: Long Papers-Volume
  1}}. Association for Computational Linguistics, \bibinfo{pages}{1035--1044}.
\newblock


\bibitem[\protect\citeauthoryear{Ratinov and Roth}{Ratinov and Roth}{2009}]%
        {ratinov2009design}
\bibfield{author}{\bibinfo{person}{Lev Ratinov} {and} \bibinfo{person}{Dan
  Roth}.} \bibinfo{year}{2009}\natexlab{}.
\newblock \showarticletitle{Design challenges and misconceptions in named
  entity recognition}. In \bibinfo{booktitle}{{\em Proceedings of the
  Thirteenth Conference on Computational Natural Language Learning}}.
  Association for Computational Linguistics, \bibinfo{pages}{147--155}.
\newblock


\bibitem[\protect\citeauthoryear{Ritter, Clark, Etzioni, et~al\mbox{.}}{Ritter
  et~al\mbox{.}}{2011}]%
        {ritter2011named}
\bibfield{author}{\bibinfo{person}{Alan Ritter}, \bibinfo{person}{Sam Clark},
  \bibinfo{person}{Oren Etzioni}, {et~al\mbox{.}}}
  \bibinfo{year}{2011}\natexlab{}.
\newblock \showarticletitle{Named entity recognition in tweets: an experimental
  study}. In \bibinfo{booktitle}{{\em Proceedings of the conference on
  empirical methods in natural language processing}}. Association for
  Computational Linguistics, \bibinfo{pages}{1524--1534}.
\newblock


\bibitem[\protect\citeauthoryear{Tilk and Alum{\"a}e}{Tilk and
  Alum{\"a}e}{2016}]%
        {tilk2016bidirectional}
\bibfield{author}{\bibinfo{person}{Ottokar Tilk} {and} \bibinfo{person}{Tanel
  Alum{\"a}e}.} \bibinfo{year}{2016}\natexlab{}.
\newblock \showarticletitle{Bidirectional Recurrent Neural Network with
  Attention Mechanism for Punctuation Restoration.}. In
  \bibinfo{booktitle}{{\em Interspeech}}. \bibinfo{pages}{3047--3051}.
\newblock


\bibitem[\protect\citeauthoryear{Zhang}{Zhang}{2017}]%
        {Zhang}
\bibfield{author}{\bibinfo{person}{Anthony Zhang}.}
  \bibinfo{year}{2017}\natexlab{}.
\newblock \bibinfo{title}{Speech Recognition}.
\newblock   (\bibinfo{year}{2017}).
\newblock
\showURL{%
\url{https://github.com/Uberi/speech_recognition#readme}}


\end{thebibliography}

\end{document}